\documentclass[%
 reprint,
 amsmath,amssymb,
 aps,
]{revtex4-1}

\usepackage{graphicx}
\usepackage{dcolumn}
\usepackage{bm}
\usepackage{color}

\begin{document}


\title{A learning scheme to predict atomic forces and accelerate materials simulations}

\author{V. Botu} 
\affiliation{Department of Chemical and Biomolecular Engineering,\
			University of Connecticut, Storrs, CT 06269}
			
\author{R. Ramprasad}
\affiliation{Department of Materials Science and Engineering, 
             University of Connecticut, Storrs, CT 06269}

\begin{abstract}
The behavior of an atom in a molecule, liquid or solid is governed by the force it experiences. If the dependence of this vectorial force on the atomic chemical environment can be \emph{learned} efficiently with high-fidelity from benchmark reference results---using ``big data" techniques, i.e., without resorting to actual functional forms---then this capability can be harnessed to enormously speed up \emph{in silico} materials simulations. The present contribution provides several examples of how such a \emph{force} field for Al can be used to go far beyond the length-scale and time-scale regimes accessible presently using quantum mechanical methods. It is argued that pathways are available to systematically and continuously improve the predictive capability of such a learned force field in an adaptive manner, and that this concept can be generalized to include multiple elements.

\end{abstract}

\maketitle

The dynamic behavior of an atom in a molecule, liquid or solid is directly determined by the \emph{local} force it experiences. Nevertheless, as already pointed out by Feynman \cite{Feynman_1}, forces are generally viewed as secondary computed quantities and are obtained through the agency of the total potential energy---a \emph{global} property of the entire system. In practice, forces on atoms are obtained either as by-products during a potential energy evaluation, or from the first derivative of the potential energy with respect to the atomic positions. This outlook still permeates modern materials simulation efforts, regardless of whether one adopts a quantum mechanical or (semi-)empirical prescription for energy and force predictions. Direct and rapid access to atomic forces, given just the atomic configuration of a system (molecule, liquid, or solid), immediately makes it possible to perform efficient geometry optimizations, molecular dynamics (MD) simulations, and a host of other related and relevant simulations. If the capability to predict forces preserves the fidelity of high-level quantum mechanics based methods, but comes at a minuscule fraction of the cost, and if this capability can be extended systematically and progressively to potentially all configurational and chemical environments that an atom may experience, we will have a powerful and adaptive materials simulation scheme.  

\begin{figure*}
\centering
\includegraphics[width=7in]{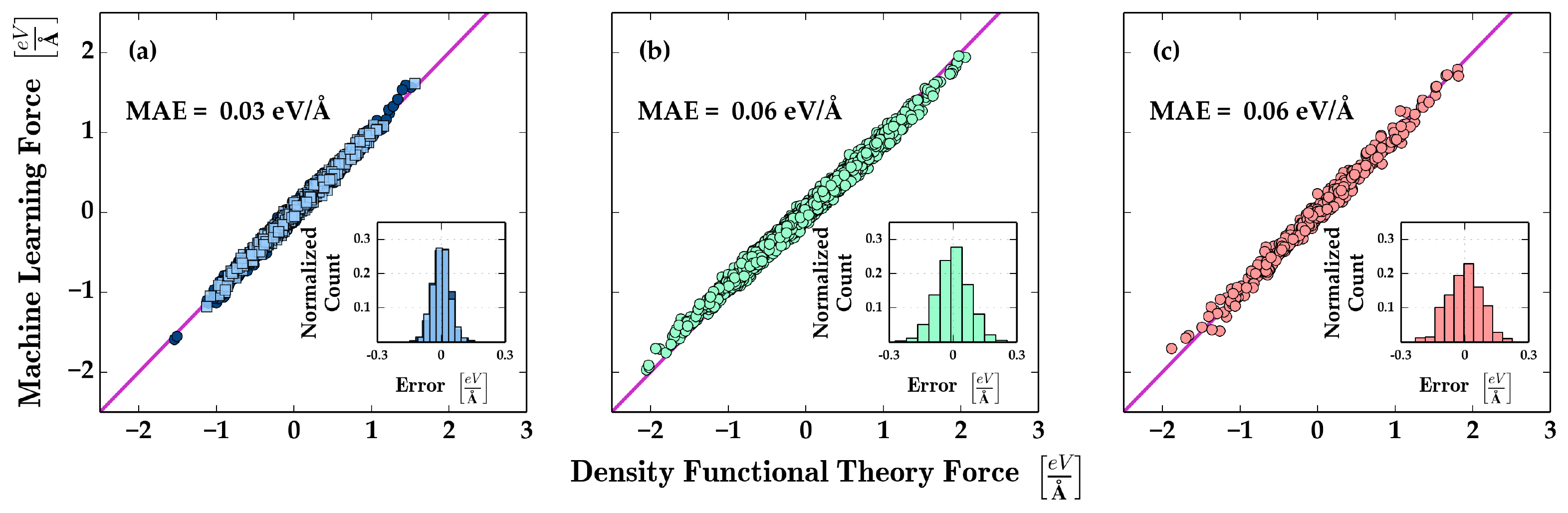}
\caption[Figure1]{Comparison of the forces predicted using the ML force field with reference DFT results, for (a) the trained model (light blue) and the validation dataset (dark blue), (b) a test unit cell containing over 800 Al atoms in the fcc phase, and (c) a test unit cell containing over 160 atoms in a hypothetical bcc phase. In (b) and (c), atoms were randomly perturbed from their equilibrium positions. Insets show the distribution of the prediction errors leading to respectable mean absolute errors (MAE).} \label{Figure 1}
\end{figure*}

The present contribution lays the ground work and takes initial steps towards the above vision. A new scheme is presented that systematically \emph{learns} in an interpolative manner to predict atomic forces in environments encountered during the dynamical evolution of materials from a set of reference atomic configurations and high-level calculations. This concept is resonant with emerging data-driven (or ``big data" \cite{Ginsberg_1, Choi_1, Hampton_1,Gobble_1,Metaxas_1}) approaches aimed at materials discovery in general \cite{Rampi_1, Ghiringhelli_1}, as well as at accelerating materials simulations \cite{Behler_2, Lorenz_1, Bartok_2, Botu_1, Li_1}. Machine learning (ML) methods using neural networks \cite{Behler_2, Lorenz_1} and Gaussian processes \cite{Botu_1, Bartok_2} have been successful in the development of interatomic potentials, wherein the potential energy surface is learned from a set of higher-level (quantum mechanics based) reference calculations.

The distinctive aspect of the present contribution, namely, learning to predict atomic forces directly (rather than the potential energy) from past data is far more powerful, but has been suggested only recently \cite{Botu_1,Li_1} (to accelerate \emph{ab initio} MD simulations on-the-fly). Here, we propose the creation of a stand-alone purely data-driven force prediction recipe devoid of any explicit functional form. This \emph{force field} is adaptive (i.e., new congurational environments can be systematically incorporated as required), generalizable (i.e., the scheme can be extended to any collection of elements for which reliable reference calculations can be performed), accurate (i.e., forces can be predicted to within  0.05 eV/$\textrm{\AA}$ of the reference calculations) and fast (i.e., a speed-up of over 8 orders of magnitude with respect to the corresponding quantum mechanics based simulations may be expected). A practical scheme that exploits the rapid high-fidelity force prediction capability within a materials simulation framework is presented, and demonstrated for Al in several configurational environments and dynamical situations that go well beyond the reaches of conventional first principles simulations. Pathways to extend this concept to handle multi-elemental systems are also proposed.

Central to this  development is a robust scheme to numerically and simply represent, or \emph{fingerprint}, the atomic environments. Such a fingerprint should differentiate dissimilar configurations with adequate accuracy, and be invariant to transformations of the environment such as translation, rotation and permutation of like elements. While several such prescriptions have been proposed in the past \cite{Botu_1, Li_1, Behler_1, Yang_1, Bartok_1, Rupp_1, Pilania_1}, the present objective, namely, mapping the vectorial force experienced by an atom to its configurational environment, places stringent constraints on the nature of the fingerprint. We argue that the following fingerprint function, $V_i^k(\eta)$, may be used to accurately represent the $k^{\textrm{th}}$ component of the force on atom $i$:
\begin{equation}\label{eq:atom}
V_i^k(\eta) = \sum\limits_{j \neq i} \frac{r_{ij}^k}{r_{ij}} \cdot e^{-{\left(\frac{r_{ij}}{\eta}\right)}^2} \cdot f{(r_{ij})}.
\end{equation}
$r_{ij}$ is the distance between atoms $i$ and $j$, while $r_{ij}^k$ is a scalar projection of this distance along component $k$. To determine the force on an atom, we require three such components along non-parallel directions. The parameter $\eta$ governs the extent of co-ordination around atom $i$  that needs to be captured. The fingerprint is essentially a spectrum of $V_i^k$ values corresponding to predetermined choices of $\eta$ values, i.e., $V_i^k$ is defined in an $\eta$-grid. The diminishing influence of faraway atoms is handled by a damping function, $f(r_{ij}) = 0.5\left[\cos\left(\frac{\pi r_{ij}}{R_c}\right)+1\right]$. The summation in Eq. \ref{eq:atom} runs over all neighboring atoms within an arbitrarily large cutoff distance $R_c$ (8 \AA, in the present work). By construction, the fingerprint will lead to symmetry-adapted forces. For instance, an atom in a centro-symmetric position will lead to a fingerprint with all zero values (and should correspond to a zero force).

The next step is to map the fingerprints to appropriate force components. Here, we have adopted the kernel ridge regression (KRR) method, capable of handling complex non-linear relationships \cite{Rupp_1, Pilania_1, Botu_1}. The KRR method works on the principle of similarity. By comparing an atom's fingerprint, $V_i^k(\eta)$, with a set of reference cases, an interpolative prediction of the $k^{\textrm{th}}$ component of the force ($F_i^k$) can be made, and is given by   
\begin{equation}\label{eq: krr}
F_i^k = \sum\limits_{t} {\alpha_t \cdot \textrm{exp}{\left[-\frac{||V_i^k(\eta) - V_t^k(\eta)||^2}{2\sigma^2}\right]}}.
\end{equation}
$t$ labels each reference atomic environment, and $V^k_t(\eta)$ is its corresponding fingerprint. $||V_i^k(\eta) - V_t^k(\eta)||$ is the Euclidean distance between the two atomic fingerprints, though other distance metrics can be used. $\alpha_t$s and $\sigma$ are the weight coefficients and length scale parameter, respectively. The optimal values for $\alpha_t$s and $\sigma$ are determined during the training phase, with the help of cross-validation and regularization methods \cite{Botu_1, Rupp_1, Hansen_1, Hastie_1}. 

\begin{figure*}
\centering
\includegraphics[width=7in]{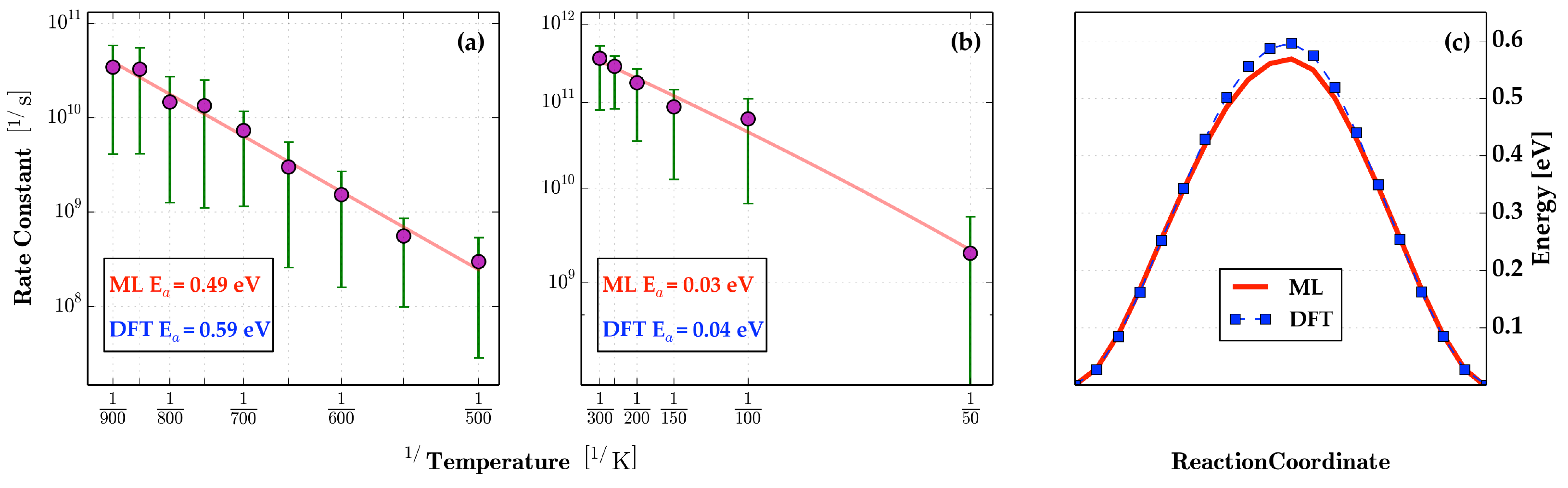}
\caption[Figure3]{Arrhenius plots for (a) vacancy migration in bulk Al and (b) adatom diffusion on the Al (111) surface. For each temperature, the MD simulation time was extended so as to allow at least 50 hopping events (thus allowing estimation of an average hop rate, and the indicated error bar). A linear fit (solid red line) was used to determine the dynamic activation energy (ML E$_a$), and is compared with the static DFT activation energy (DFT E$_a$). (c) For the vacancy migration in bulk Al, the DFT potential energy along the migration trajectory (symbols and dashed line), and the corresponding energy obtained via an integration of the ML forces along the reaction coordinate (solid line).} \label{Figure 3}
\end{figure*}  

Using the above prescribed framework, a ML force field for Al has been developed using a plethora of reference atomic environments accumulated from density functional theory (DFT) based MD runs at various temperatures using the Vienna ab initio simulation package \cite{Kresse1, Kresse2, Perdew1, Blochl1} (other means may also be used to generate the reference data, as long as they satisfy prescribed  demands on accuracy of the atomic forces). To ensure a diverse set of reference cases, Al in different geometric arrangements were considered (but each one with just a few tens of atoms per repeating unit cell), including defect-free bulk in the face centered cubic (fcc) phase, bulk fcc phase with vacancy, clean (111) surface, and the (111) surface with adatom, resulting in over 100,000 atomic environments \cite{Botu_1}. Interestingly, a random set of 1000 atomic environments drawn from the accumulated environments proved sufficient to construct an accurate interpolative force prediction model. Figure \ref{Figure 1}(a) compares the predicted forces with the DFT forces (including the error distribution in the inset) for all accumulated configurations, i.e., those used in the training phase and the remaining configurations whose results were used for validation. The mean absolute error (MAE) of the prediction model was 0.03 eV/\AA, of the order of the expected chemical and numerical accuracy of the reference DFT calculations. Needless to say, in addition to providing a pathway to accurately predict atomic forces, this procedure is also extremely expedient; it scales linearly with system size, and can be well over 8 orders of magnitude faster than a typical DFT calculation. 

An immediate (and straight-forward) application of this fast high-fidelity capability to predict atomic forces is geometry optimization, including the prediction of potential energy minima and saddle points. Simulations involving hundreds of thousands of atoms (i.e., cases that are beyond the reaches of present day DFT computations) can be handled, provided the chemical environments encountered during the course of such optimizations are included in the force field. In order to understand the limits of the constructed ML force field for Al within the context of such simulations, a few tests were performed. The first one involved a large unit cell containing over 800 Al atoms in the fcc phase along with Al vacancies. Atoms were randomly perturbed, and the ML force field was used to optimize this perturbed structure. The correct equilibrium geometry was recovered, as ascertained by a separate DFT calculation starting with the same perturbed system. A video of this optimization is included in the Supplemental Information. Figure \ref{Figure 1}(b) compares the predicted forces with the DFT forces for the initial perturbed geometry. Much larger unit cells could be considered using the ML force field but we restrict ourselves to modest sizes in this discussion as we are constrained by the inability of DFT to provide validation for truly large unit cells.

As a second geometry optimization example, a 160 atom unit cell of Al in a hypothetical body centered cubic (bcc) phase was considered. Once again, the atoms were perturbed randomly, followed by geometry optimization. Figure \ref{Figure 1}(c) captures the performance of the force field for the starting geometry. Given that the bcc phase was never used in the training phase during the force field creation, we would expect that forces on atoms in such an environment will be difficult to predict. Surprisingly, going by the rather small force error distribution (comparable to the 800-atom fcc example), we conclude that the current choice of fingerprints allows us to effectively capture diverse chemical environments in a versatile manner. 

Next, we consider non-zero temperature dynamical situations. For the force prescription to correctly capture dynamic processes with high-fidelity, ergodicity has to be preserved. In other words, the average behavior and time scales of elementary steps or processes should be correctly represented during a MD simulation using the force field. As a first example, we consider the diffusion of an Al vacancy in bulk Al, using a unit cell containing 32 Al sites and an Al vacancy. MD simulations were performed at 9 temperatures in the 500-900 K range for times up to 5 ns, with a timestep of 0.5 fs. By observing the dynamics of the vacancy, the average rate constant ($k$) for the migration process at each temperature was determined. $k$ is given as $^{1/}t_{hop}$, where $t_{hop}$ is the average time taken for a vacancy to migrate to a neighboring site. To ensure that sufficient statistics are collected, $k$ was averaged over 50 such hop events at each temperature. Figure \ref{Figure 3}(a) shows an Arrhenius plot of $k$ versus the reciprocal temperature, whose slope yields the activation energy (E$_a$) for Al vacancy migration to be 0.49 eV. The corresponding DFT value for a similar, but static, migration process was determined to be 0.59 eV (c.f., Figure \ref{Figure 3}(c)). Barrier ``softening" is expected under dynamical conditions, relative to the results of static calculations in which entropic effects are neglected \cite{Mer_1, Benson_1}.

Another elementary process we considered was the self-diffusion of an Al adatom on the Al (111) surface, using a 6 $\textrm{\AA}$ x 6 $\textrm{\AA}$ surface unit cell containing a 4-layer thick Al slab. Similar to the Al vacancy example, by monitoring the dynamics of the adatom across a temperature range of 50-300 K, an E$_a$ of 0.03 eV was predicted, as shown in Figure \ref{Figure 3}(b), whilst a static DFT calculation yielded an E$_a$ of 0.04 eV. A video of the adatom migration MD simulation is included in the Supplemental Information.

Both dynamical diffusion scenarios considered lead to the correct Arrhenius behavior indicating that the underlying physics is properly captured in the ML force field based MD simulations. Moreover, although the force field is aimed at directly predicting atomic forces, potential energy differences for elementary steps may be obtained by integrating the forces along a suitable reaction coordinate. Figure \ref{Figure 3}(c), for instance, portrays the DFT energy profile along the Al vacancy migration pathway in bulk Al, as well as the corresponding energy determined by integrating the forces predicted by the ML force field. The close agreement between the two energies is self-evident, indicating that energies corresponding to critical parts of a trajectory may indeed be obtained from the forces through integration. 

\begin{figure}[b]
\centering
\includegraphics[width=3in,height=3in]{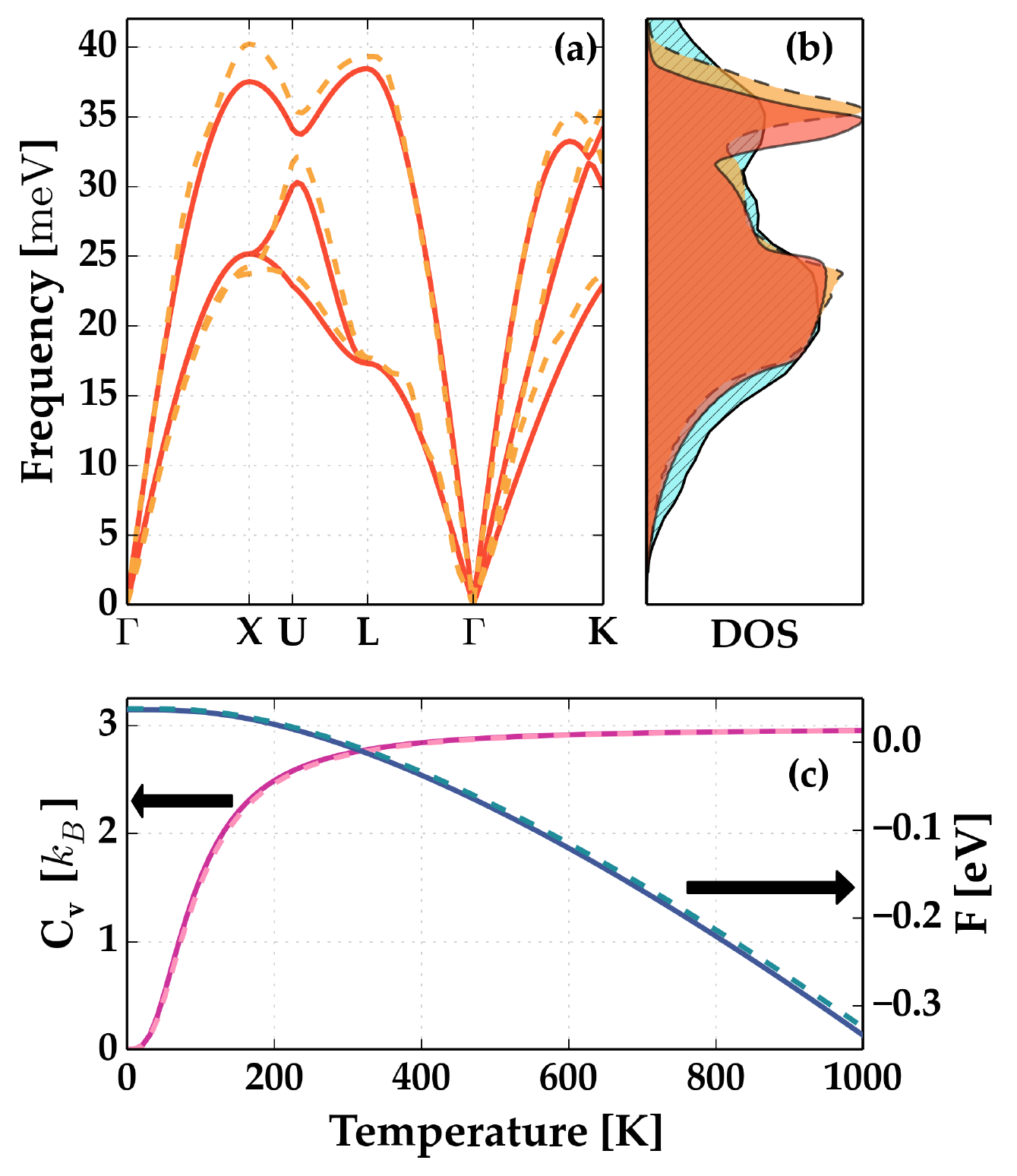}
\caption[Figure4]{(a) Phonon band structure, (b) phonon density of states (DOS), and (c) Helmholtz free energy and constant volume heat capacity computed using the ML force field (solid lines) and DFT (dashed lines). The phonon band structure and DOS were computed using the finite atomic displacement method. Also included in (b) are the DOS results obtained from the Fourier transform of the velocity autocorrelation function (solid cyan hatched fill).} \label{Figure 4}
\end{figure} 

Lastly, we evaluate the prospect of how well thermal behavior of materials can be simulated using the force-based framework. In particular, we focus on the vibrational (or phonon) density of states (DOS), which has to be properly captured to allow for accurate calculations of thermodynamic quantities, thermal expansion, thermal conductivity, etc. Figure \ref{Figure 4}(a) shows the phonon band structure as determined using the ML force field and using DFT, and in both cases, the finite displacement method was used \cite{Alfe_1}. Figure \ref{Figure 4}(b) shows the corresponding DOS, as well as the DOS computed using the Fourier transform of the velocity autocorrelation obtained from a MD simulation \cite{Dickey_1}. This latter approach implicitly includes anharmonicity to all orders (the first method, in contrast, includes just the harmonic part). The MD simulation involved a 864 atom unit cell, and a simulation time of 5 ps at 700 K. Excellent agreement of the ML force field result with the reference DFT calculations can be seen. The deviations of the DOS computed using MD simulations relative to that obtained using the finite displacement scheme (especially at high frequencies) may be attributed to non-zero anharmonic effects. The DOS can be utilized to determine thermodynamic properties such as the Helmholtz free energy and the constant volume heat capacity. These properties, as a function of temperature, are compared with the corresponding DFT results in Figure \ref{Figure 4}(c). The ML force field and DFT results are nearly indistinguishable, indicating that even under the stringent test of small atomic perturbations encountered in these situations (as opposed to the larger length scale vacancy or adatom hops discussed earlier), the fidelity of the force prediction is preserved. 

A natural question that arises at this point is how this force field paradigm may be extended to include multiple elements. In a multi-elemental system, the fingerprint of an atom of a given element type may be constructed to have as many parts as the number of elements in the system. Each part would represent the arrangement of atoms of a particular elemental type around the reference atom. While this scheme requires further optimization, preliminary work shows significant promise. For two binary systems, $\alpha$-Al$_2$O$_3$ and monoclinic HfO$_2$, the force prediction based on the concatenated multi-component fingerprint prescription rivals that for the elemental Al in quality. A parity plot comparing the predicted force with the corresponding reference DFT result for each element type is shown in the Supplemental Information. Given such accuracies, extension of the proposed concept to multielemental systems appears feasible.

The discussion thus far has provided an expos{\'e} of materials simulation examples that can benefit enormously through a capability to directly and rapidly predict atomic forces with demonstrable verisimilitude. This capability learns from past reference quantum mechanical calculations, but can access length-scales and time-scales significantly beyond the reaches of purely quantum mechanics based simulations (while preserving accuracy). Examples of phenomena that can potentially be studied include transport (thermal and mass), phase transformations and chemical reactions, mechanical behavior, materials degradation and failure, etc., all within the framework of reality-mimicking non-zero temperature dynamical simulations. Widespread use of the proposed class of learning-based \emph{force} fields will require attending to a few critical matters. These include: (i) creation of an initial compact training set of reference atomic environments appropriate for a particular materials application; and, (ii) development of a capability to recognize a truly new atomic environment when such is encountered during the course of a simulation. The latter aspect is critical to evaluating when the force field is expected to fail, and, as importantly, to supplement the initial training set so as to make the force prediction scheme adapt, evolve and continuously improve with time. Nevertheless, these hurdles have been encountered, and addressed, in the past in many ``big data" situations \cite{Ginsberg_1, Choi_1, Hampton_1, Gobble_1, Metaxas_1}. Hence, there is reason for (cautious) optimism in the present context of high-fidelity, adaptive and generalizable atomic force fields.


This work was supported financially by a grant from the Office of Naval Research (N00014-14-1-0098). The authors would like to acknowledge helpful discussions with K. B. Lipkowitz, G. Pilania, T. D. Huan, A. Mannodi-Kanakkithodi and A. Dongare. Partial computational support through a Extreme Science and Engineering Discovery Environment (XSEDE) allocation is also acknowledged. All calculations were performed in the pythonic environment, with the atomic simulation environment, pwtools and mlpy modules \cite{ASE,pwtools,mlpy}.


%
\end{document}